\documentclass{article}

\usepackage[letterpaper, total={6in, 8in}]{geometry}
\usepackage{amssymb,amsfonts,amsmath}
\usepackage{graphicx}
\usepackage{xcolor}
\usepackage{hhline}
\usepackage{multirow}
\usepackage{array}
\usepackage{booktabs}
\usepackage{siunitx}
\usepackage{url}
\usepackage{authblk}

\title{A method to estimate the rectangular orthotropic plate elastic constants using least-squares and Chladni patterns}

\author[1]{Michele Ducceschi}
\author[1]{Sebastian Duran}
\author[1]{Henna Tahvanainen}
\author[2]{Ludovico Ausiello}
\affil[1]{Department of Industrial Engineering, University of Bologna, Italy}
\affil[2]{SENE, University of Portsmouth, UK}

\begin{document}

\maketitle

\begin{abstract}
A method to retrieve the elastic constants of rectangular wooden plates is presented, relying on the measurement of a set of eigenfrequencies and the identification of the corresponding mode shapes, and belonging to the more general category of non-destructive inverse parameter estimation methods. Compared to previous works, the current method is effective with any choice of boundary conditions. Furthermore, the error function is linear in the elastic constants, which are computed via a matrix inversion. This framework lends itself naturally to a physical interpretation of the results in terms of linear combinations of eigenmodes, yielding new sets of modes and associated combined mode shapes in which the elastic constants are completely uncoupled.
A number of numerical benchmark tests and experimental cases are treated in detail, highlighting the reliability of the proposed methodology in cases of interest in acoustics and musical acoustics.
\end{abstract}

\section{Introduction}
\label{sec:Introduction}
.

Inverse parameter estimation refers to a wealth of techniques used to identify model parameters in materials such as wood \cite{mcintyre_AM_1988,caldersmith_JCA_1990}, composites \cite{viala_COMP_2018,ege_JSV_2018}, ceramics \cite{jin_EFM_2014,jin_EFA_2016}, metals \cite{maciolek_IJF_2021,chen_JMPT_2016} and others. Commonly, one looks for parameters minimising the discrepancy between the experimental data and a suitable reference model, using various techniques differing in application and methodology \cite{tam2017identification}. A typical distinction in this sense considers destructive and non-destructive techniques. The measurement of the elasticity constants through traction, compression, bending and twisting is a typical example of the former: for elastic specimens, some form of permanent damage usually results when the applied forces operate beyond the linear regime \cite{chen_JMPT_2016,bruhwiler_EFM_1990}. Dynamic or vibrational testing, on the other hand, is an example of the latter. In this case, the operating regime is most often linear, and the measurement setup does not usually produce damage to the specimens \cite{mcintyre_AM_1988,caldersmith_JCA_1990,viala_COMP_2018,ege_JSV_2018}. 

Inverse parameter estimation is useful to measure the parameter variability across wood samples, the subject of this work. Wood is often modelled as an orthotropic material, depending on nine independent elastic constants: three Young's moduli, three shear moduli, and three Poisson's ratios \cite{bucur_book_2016}. Inverse parameter techniques have been successfully devised for this case, for example employing 3D laser vibrometry and finite element model updating \cite{viala_COMP_2018,mottershead_MSSP_2011}. Differences across cuts from the same tree have been reported \cite{vanCasteren_PLOS_2013}, \cite[ch.3]{Bootle_wood}, and \cite[ch.4]{gore2011contemporary}. Furthermore, wood is susceptible to changes in environmental conditions such as temperature and moisture \cite{you_SR_2021}, as well as to chemical treatment \cite{albano_HS_2022,malvermi_SR_2022}. These changes entail macroscopic modifications in a sample's density and size, as well as in its response to excitation \cite{persson_AA_2019}. Whilst the former are easily recognised and measurable, the latter is established by changes in the sample's elastic properties, which are harder to quantify.

In some cases, a reduction of the number of elastic constants is possible through further approximations, such as when the sample is thin, a case most often encountered in musical instruments where typical thicknesses are found in the range of a few millimetres \cite{fletcher_book_2012}. Well-established techniques for obtaining the elastic constants of thin rectangular plates exist, such as the ones presented by McIntyre and Woodhouse \cite{mcintyre_AM_1988} and by Caldersmith and Freeman \cite{caldersmith_JCA_1990}. In these cases, the elastic constants are retrieved by measuring the frequencies of three specific modes of a plate with free edges, and by comparing them with approximate analytic solutions obtained via the Rayleigh method. These techniques have a simple and inexpensive setup and have become commonplace among scientists and instrument builders alike \cite{bucur2006acoustics, malvermi_phd,quintavalla2022grading}. 

In this work, an extension of such techniques is offered, not relying on a specific set of boundary conditions or a fixed set of modes, and leveraging numerical simulation. It will be shown that, in the thin-plate approximation, the modal frequencies depend linearly on the elastic constants, where the linear coefficients can be extracted from a batch of numerical training plates sharing the same boundary conditions and aspect ratio as the experimental plate. 

Once such modal coefficients are known, the elastic constants can be estimated immediately via a matrix inversion in a least-square sense. Compared to well-established techniques in the literature, the current method offers the possibility to perform multiple estimates of the elastic constants, using various combinations of boundary conditions and modes. Furthermore, the mathematical problem is linear, thus avoiding the need for nonlinear solvers or iterative methods for which the existence and uniqueness of the solution may not be guaranteed \cite{burke_MP_1995}. One only needs to associate the experimental frequency to a matching set of modal coefficients depending on the nodal lines on the plate's surface, which can be achieved quickly using the known method by Chladni \cite{rossing1982chladni,chladni2015treatise}.
Numerical benchmark tests and experimental results show the applicability of the method to the measurement of the elastic constants of spruce boards. The paper is structured as follows: Section \ref{sec:Methods} presents the methodology, with a specific focus on the linear dependence of the non-dimensional squared plate frequencies on the elastic constants. Section \ref{sec:NumBench} presents a few numerical benchmark tests, highlighting the feasibility of the proposed methodology under two different choices of boundary conditions. Section \ref{sec:experiments} describes the experimental setup to obtain the frequencies of a cantilever Finnish spruce plate and a clamped red spruce plate. Finally, Section \ref{sec:Discussion} presents the estimates of the elastic constants for the two experimental plates and a discussion on the validity of the method.

\section{Methodology}\label{sec:Methods}

The flexural vibration of a thin, rectangular wood panel can be described by the orthotropic Kirchhoff plate equation \cite{Aksu_PhD}. This is:
\begin{equation}\label{eq:DimEqCnt}
\rho L_z \, {\partial_t^2 u} = -\left( D_{x}  ({\partial_x^4 } + \nu_{yx} {\partial_x^2\partial_y^2 }) +  D_{y} ({\partial_y^4 } + \nu_{xy} {\partial_x^2\partial_y^2 }) + D_s \, {\partial_x^2\partial_y^2 } \right) u.
\end{equation}
Here, $u = u({\bf x},t): \mathcal{V} \times \mathbb{R}^+_0 \rightarrow \mathbb{R}$ is the flexural displacement of the plate, a function of the spatial coordinates ${\bf x} \in \mathcal{V} = \{(x,y) \, | \, 0 \leq x \leq L_x, \, 0 \leq y \leq L_y\}$, as well as time $t \geq 0$. In the above, $\partial_r^l$ denotes the $l^{th}$ partial derivative with respect to $r$. Boundary conditions of classic type are imposed at the plate's edges, such as free, clamped or simply-supported. Let $\mathcal{B}$ denote the set of boundary conditions imposed along the plate's edges, such that, for example, $\mathcal{B} = \{\text{C-S-F-C}\}$ denotes a plate with a clamped, a simply-supported, a free and another clamped edge. For an edge perpendicular to the $y$ direction, these are defined as follows \cite{Aksu_PhD}:
\begin{subequations}
\begin{align}
\text{C: }& u = \partial_x u = 0, \\
\text{S: }& u = (\partial^2_x + \nu_{yx}  \partial^2_y) u = 0, \\
\text{F: }& (D_x \partial^3_x + (D_x\nu_{yx} + D_s)\partial_x\partial^2_y)u =  (\partial^2_x + \nu_{yx}  \partial^2_y) u = 0,
\end{align}
\end{subequations}
with analogous definitions holding for an edge perpendicular to the $x$ direction (one only needs to swap $x$ and $y$ in the above). A further condition must be imposed at a corner of two free edges, namely $u_{xy} = 0$ (this is known as the Kirchhoff corner condition).

In the model above, constants appear as: $\rho$, the volume density of the plate (in kg $\cdot$ m$^{-3}$); the plate dimensions $L_x,L_y,L_z$ (in m, and where the thickness $L_z$ is assumed constant across $\mathcal{V}$); the three rigidity constants $D_x$, $D_y$, $D_s$ (in N $\cdot$ m). These are defined by:
\begin{equation}\label{eq:RigCnst}
D_x = \frac{E_xL_z^3}{12(1-\nu_{xy}\nu_{yx})}, \qquad
D_y = \frac{E_yL_z^3}{12(1-\nu_{xy}\nu_{yx})}, \qquad
D_s = \frac{G_{xy}L_z^3}{3},
\end{equation}
where $E_{x}$, $E_y$ are  Young's moduli (in Pa), $\nu_{xy}$, $\nu_{yx}$ are  dimensionless Poisson's ratios, $G_{xy}$ is a shear modulus (in Pa). The rigidity constants are defined in terms of five elastic constants ($E_{x}$, $E_y$, $G_{xy}$, $\nu_{xy}$, $\nu_{yx}$). However, due to the symmetry of the compliance matrix in the elasticity equations, only four of them are independent \cite{wilczynski2011determination}. Usually, one of the two Poisson's ratios is set according to:
\begin{equation}\label{eq:nuyx}
E_x \nu_{yx} = E_y\nu_{xy}.
\end{equation}

\begin{figure} 
\centering
\centering
    \includegraphics[width = \linewidth, clip, trim = 1cm 3cm 1cm 1cm ]{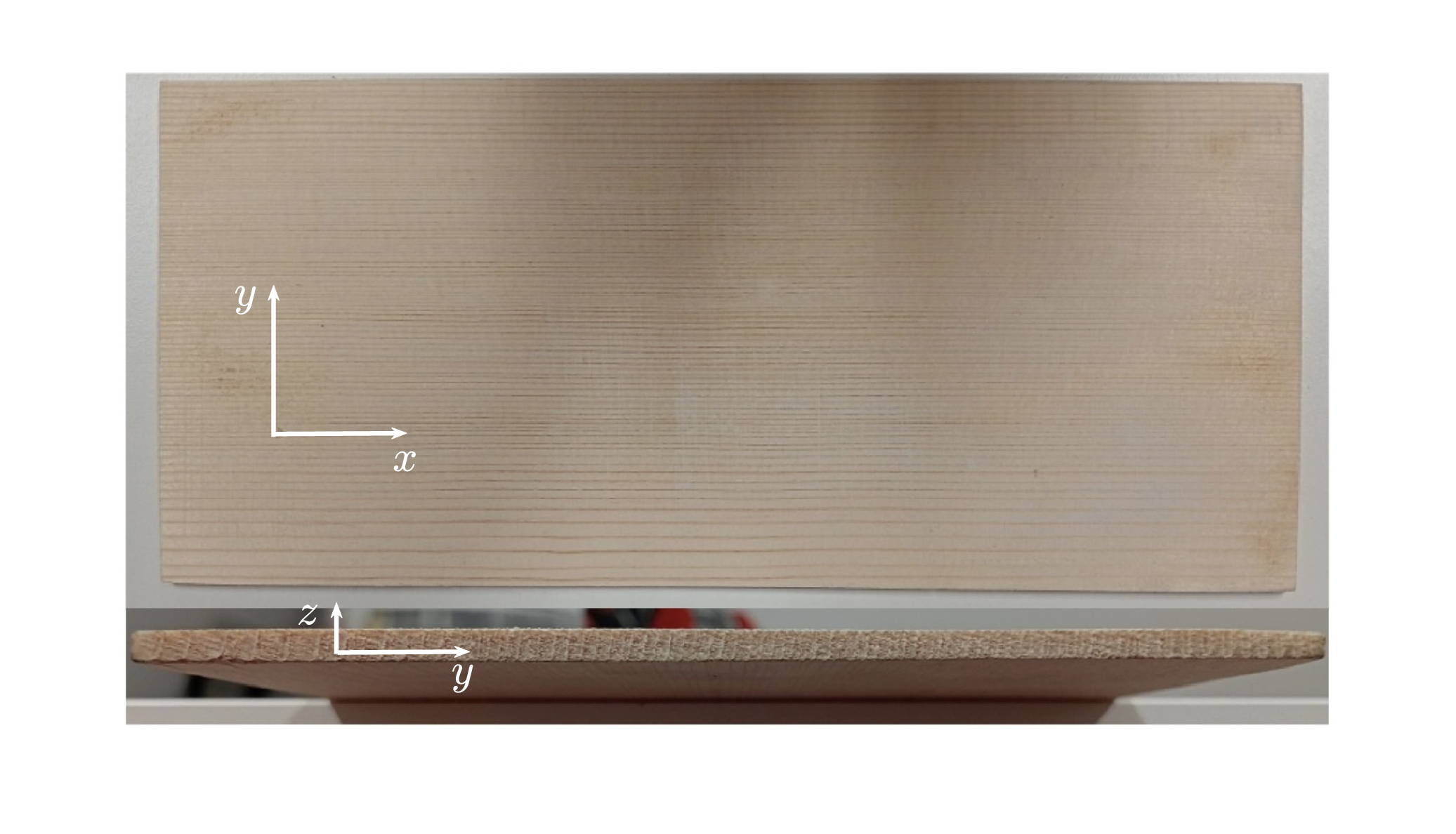}
\caption{Fiber direction convention in this article. $x$: longitudinal; $y$: radial; $z$: tangential. } \label{fig:Tonewood} 
\end{figure}

The wood's fibre orientation is responsible for the orthotropic character of lumber and results from the particular cutting technique employed. In quarter sawing, logs are quartered lengthwise with the annual rings placed almost perpendicular to the board faces \cite{how2007review}. This technique, whilst wasteful, produces boards with an increased resistance against environmental factors such as moisture and is most prized in musical instrument making. Conveniently, the axes' orientation may be referred back to the fibres. In the following, $x$ will denote the direction along the grain (longitudinal); $y$ will denote the direction across the grain (radial); and $z$ will denote the tangential direction; see also Figure \ref{fig:Tonewood}. The stiffness in the longitudinal direction is usually an order of magnitude higher than in the other two directions, resulting in increased wave propagation velocity at reference wavelengths. 

\subsection{Non-dimensional model}\label{sec:NonDimModel}

Equation \eqref{eq:DimEqCnt} depends on six constants: $\rho$, $L_z$, $E_x$, $E_y$, $G_{xy}$, $\nu_{xy}$, as well as the side lengths $L_x, L_y$. Inverting a system with such a large number of parameters may be impractical and lead to non-uniqueness of the resulting solution, an issue often emerging in inverse parameter estimation methods \cite{berg_CG_2021}. In some cases, typically in machine learning and, particularly, deep learning frameworks, interpretability is unnecessary so long as the expected system's output is produced \cite{esqueda2021differentiable}. In other cases, such as here, a physical interpretation of the model may be obtained by recasting the problem in a form needing a minimal amount of independent parameters, as one may do immediately through non-dimensionalisation. To that end, divide both sides of \eqref{eq:DimEqCnt} by $\rho L_z$, and collect $D_{s}$ on the right-hand side. This gives:
\begin{equation}\label{eq:SemiDim}
{\partial_t^2 u} = -\kappa^2 \left(  ({\partial_x^4 }+\nu_{yx}\partial_x^2\partial_y^2)\,p +  ({\partial_y^4 }+\nu_{yx}\partial_x^2\partial_y^2)\,q +  \, {\partial_x^2\partial_y^2 } \right) u,
\end{equation}
where $\kappa^2 := \rho^{-1} L_z^{-1}D_{s}$, $p :=  D^{-1}_{s} D_x$, $q :=  D^{-1}_{s} D_y$. Non-dimensionalisation of the system proceeds as follows. Define:
\begin{equation}\label{eq:NDimDef}
u = {A}^{\tfrac{1}{2}} \,  \bar u, \,\, x = {A}^{\tfrac{1}{2}}  \,  \bar x, \,\, y = {A}^{\tfrac{1}{2}}  \, \bar y, \,\, t = \kappa^{-1}{A \bar t},
\end{equation}
where the overbar notation denotes a non-dimensional variable, and where $A := L_xL_y$ is the plate surface area. Thus, \eqref{eq:SemiDim} becomes:
\begin{equation}\label{eq:NonDim}
{\partial_{\bar t}^2 \bar u} = - \left(  ({\partial_{\bar x}^4 }+\nu_{yx}\partial_{\bar x}^2\partial_{\bar y}^2)\,p +  ({\partial_{\bar y}^4 }+\nu_{yx}\partial_{\bar x}^2\partial_{\bar y}^2)\,q +  \, {\partial_{\bar x}^2\partial_{\bar y}^2 } \right) \bar u,
\end{equation}
defined over the non-dimensional domain $\mathcal{\bar V} = \{(\bar x,\bar y) \, | \, 0 \leq \bar x \leq \sigma^{\tfrac{1}{2}}, \, 0 \leq \bar y \leq \sigma^{-\tfrac{1}{2}}\}$, with $\sigma:=L_y^{-1}L_x$ denoting the aspect ratio of the plate. Equation \eqref{eq:NonDim} depends only on the two ratios  $p, q$, as well as $\sigma$.

Transforming \eqref{eq:NonDim} in the frequency domain (that is, using $\partial_{\bar t} \rightarrow j\bar \omega$, $\partial_{\bar x} \rightarrow j\bar \gamma_{\bar x}$,  $\partial_{\bar y} \rightarrow j\bar \gamma_{\bar y}$ for temporal frequency $\bar \omega$ and wavenumbers $\bar \gamma_{\bar x}$, $\bar \gamma_{\bar y}$), the non-dimensional dispersion relation is recovered:
\begin{equation}\label{eq:NDimDispRel}
\bar \omega^2 =  (\bar \gamma_{\bar x}^4 + \nu_{yx} \bar \gamma_{\bar x}^2\bar \gamma_{\bar y}^2)\, p +  (\bar \gamma_{\bar y}^4 + \nu_{xy} \bar \gamma_{\bar x}^2\bar \gamma_{\bar y}^2)\, q  + \bar \gamma_{\bar x}^2 \bar \gamma_{\bar y}^2,
\end{equation}
from which the modal frequencies $\bar \omega_{m,n}$,  are extracted as a quantised version of \eqref{eq:NDimDispRel}. Here, $m,n \in \mathbb{N}$ denote a pair of modal indices related to the nodal lines in the $x$ and $y$ directions. For fixed $p$, $q$, the quantisation depends exclusively on the boundary conditions $\mathcal{B}$ and the aspect ratio $\sigma$. On the other hand, when one allows $p$, $q$ to vary  keeping $\sigma$ and $\mathcal{B}$ fixed, a linear dependence is assumed:
\begin{equation}\label{eq:linDepOm}
\bar\omega^2_{m,n} = a_{m,n} \, p + b_{m,n} \,  q + c_{m,n},
\end{equation}
for modal number-dependent coefficients $(a_{m,n},b_{m,n},c_{m,n}) \geq 0$. This is a key feature of the current formulation, allowing to estimate the rigidity constants using a simple matrix inversion, as will be shown in the following sections. Before proceeding, notice that
the dimensional modal frequencies $\omega_{m,n}$ are obtained via multiplication of the non-dimensional frequencies by the time scaling constant, as per \eqref{eq:NDimDef}:
\begin{equation}\label{eq:OmegaDef}
\omega^2_{m,n} := \kappa^2 A^{-2} \bar\omega^2_m =  D_{s} \, \rho^{-1} L_z^{-1} A^{-2} \bar \omega^2_{m,n}.
\end{equation}
Substituting the linear relationship \eqref{eq:linDepOm} in \eqref{eq:OmegaDef}, one obtains:
\begin{equation}\label{eq:LinDimOms}
\omega^2_{m,n} = \lambda\left(a_{m,n}D_x + b_{m,n}D_y + c_{m,n}D_{s} \right),
\end{equation}
with $\lambda := \rho^{-1} L_z^{-1} A^{-2}$.

\subsection{Problem Formulation}

In the following, assume one wishes to estimate the rigidity constants $D_x$, $D_y$ $D_{s}$ for the experimental plate under study. 
Furthermore, the plate parameters $\rho$,  $L_x,L_y,L_z$ are assumed to be known, along with the boundary conditions $\mathcal{B}$.

The method relies on minimising a suitably defined error between the dimensional numerical eigenfrequencies $\omega_{m,n}$ in \eqref{eq:LinDimOms} and a set of experimentally measured frequencies $\hat \omega_{m,n}$, where the minimisation is performed with respect to the unknown parameters $d_x,d_y,d_{s}$ (these are approximations of the ``true'' rigidity constants $D_x$, $D_y$, $D_{s}$). Thus, define:
\begin{equation}
{\boldsymbol \psi} := \lambda^{-1} \hat{\boldsymbol \omega}^2, \quad {\boldsymbol \phi}\left(d_x,d_y,d_{s} \right) :=  \, \begin{bmatrix}{\bf a}, {\bf b}, {\bf c}\end{bmatrix}\begin{bmatrix} d_x \\ d_y \\ d_{s}\end{bmatrix} := {\bf X}\,{\bf d},
\end{equation}
where ${\bf X :=   \begin{bmatrix}{\bf a}, {\bf b}, {\bf c}\end{bmatrix}}$, ${\bf d := \begin{bmatrix} d_x , d_y , d_{s}\end{bmatrix}^\intercal}$. Here, $\bf a$ is a column vector obtained by stacking $a_{m,n}$ for a set of modes. As an example, assume to work with modes $(0,1),(1,1),(1,2)$. Then, ${\bf a} = [a_{0,1},a_{1,1},a_{1,2}]^\intercal$. Analogous definitions hold for $\bf b$, $\bf c$. Thus, $\boldsymbol \phi$ is a vector expression of \eqref{eq:LinDimOms} using the approximate rigidity constants $\bf d$.

The error between the measured and the numerical frequencies is here defined as:
\begin{equation}\label{eq:EspDef}
\varepsilon\left(d_x,d_y,d_{s} \right) := \frac{1}{2} \|{\boldsymbol \psi} - {\boldsymbol \phi}\|^2_2,    
\end{equation}
from which the rigidity constant vector $\bf d$ is obtained through a least-square minimisation \cite{boyd2004convex}: 
\begin{equation}\label{eq:dEstim}
{\bf d} = \arg\min \varepsilon = \left({\bf X}^\intercal {\bf X}\right)^{-1} \left( {\bf X}^\intercal {\boldsymbol \psi}\right).
\end{equation}
From here, one computes:
\begin{align}\label{eq:EstElCnt}
e_x = \frac{12(1-\nu_{xy}\nu_{yx})d_x}{L_z^3}, \qquad 
e_y = \frac{12(1-\nu_{xy}\nu_{yx})d_y}{L_z^3}, \qquad
g_{xy} = \frac{3d_{s}}{L_z^3},
\end{align}
approximating $E_x$, $E_y$, $G_{xy}$. Given their small variability across specimens, it will be assumed that the Poisson ratios $\nu_{xy}$, $\nu_{yx}$ are known and obtained from tabulated values.

Note that the methodology relies on the knowledge of the modal coefficients  $a_{m,n},b_{m,n},c_{m,n}$. These are, generally, unknown, but they can be extracted by fitting the numerical eigenfrequencies of a number of training plates, as shown in the next section.

\section{Numerical Benchmark Tests}\label{sec:NumBench}

The methodology described above is now tested in a few numerical case studies. First, the validity of the linear relationship \eqref{eq:linDepOm} is assessed; second, the ability of the current formulation to predict the correct rigidity constant values from a batch of numerical test plates is shown.

\subsection{Linear dependence of the non-dimensional frequencies on $p,q$}\label{sec:LinDepCheck}

A key assumption in the current formulation is the linear dependence of the non-dimensional frequencies $\bar \omega^2_{m,n}$ on the elastic constant ratios $p,q$, as per \eqref{eq:linDepOm}. 
 This is now checked by computing the numerical eigenfrequencies of a set of plates created by varying reference plate parameters. All throughout, the aspect ratio and the boundary conditions are kept fixed: $\sigma = 223/114$, $\mathcal{B} = \{\text{C-F-F-F}\}$, with the clamped edge lying along $y$ (the radial direction).  The finite element simulation software package COMSOL is used for the numerical simulations. The \emph{Plate} COMSOL component is selected. Note that the input COMSOL parameters are dimensional and include the thick-plate parameters in addition to the eight parameters listed in Section \ref{sec:NonDimModel}, though the former have a small influence on the first few eigenmodes for thin and moderately thick plates. Table \ref{tab:TrainingPlateParams} summarises the dimensional parameters used in the simulations. 
\begin{table}
\addtolength{\tabcolsep}{5pt}
\centering
\begin{tabular}{cccccccc} 
\multicolumn{8}{c}{\footnotesize{Kirchhoff plate parameters}} \\
\midrule
    $\rho$  & $L_x$  & $L_y$  & $L_z$  & $E^0_x$  & $E^0_y$  & $G_{xy}$  & $\nu_{xy}$  \\ 
    \midrule
    473.9  & 22.3 & 11.4 & 3 & 10.7 & 0.716 & 0.50 & 0.51  \\
       \footnotesize{kg $\cdot$ m$^{-3}$} & \footnotesize{cm} & \footnotesize{cm} & \footnotesize{mm} & \footnotesize{GPa} & \footnotesize{GPa} & \footnotesize{GPa} & \\
    \midrule  \\
    \end{tabular}
    \begin{tabular}{ccccc} 
    \multicolumn{5}{c}{\footnotesize{Thick-plate parameters}} \\
    \midrule
    $E_z$ & $G_{yz}$  & $G_{xz}$   & $\nu_{yz}$  & $\nu_{xz}$  \\ 
    \midrule 
     0.39  & 0.023 & 0.62 & 0.45 & 0.50    \\
     \footnotesize{GPa} &  \footnotesize{GPa} &  \footnotesize{GPa} &  &  \\    \bottomrule
\end{tabular}
\caption{Input constants for the COMSOL simulations, inspired from the table in \cite{bucur_book_2016} (page 96). These constants are used to create a ``training'' set from which the modal coefficients $a_{m,n}$, $b_{m,n}, c_{m,n}$ are estimated. The plates included in the training set have material and geometrical constants as per the table, and where $(E_x,E_y)$ are selected as the twenty-five elements of the set $\mathcal{T} = \{E_x,E_y \, | \, E_x = (0.8,0.9,1.0,1.1,1.2)E_x^0, \, E_y = (0.8,0.9,1.0,1.1,1.2)E_y^0\}$.}\label{tab:TrainingPlateParams}
\end{table}
From the table, twenty-five combinations of the elastic constant ratios $p,q$ are created, with $4.3 < p < 6.6$, $0.28 < q < 0.45$. The modal frequencies computed in COMSOL are then sorted according to mode number and belong to the modes (1,0), (1,1), (2,0), (1,2), (2,1), (2,2).  In order to assign the correct modal frequency to each modal shape when batch exporting from COMSOL, the modal assurance criterion (MAC) can be used. Let ${\bf U}, {\bf U}^\prime$ denote the column vectors containing the sampled modal displacements of two modes. The MAC index, a scalar value, is defined as \cite{pastor2012modal}:
\begin{equation}
\text{MAC}({\bf U}, {\bf U}^\prime) := \frac{|{\bf U}^\intercal {\bf U}^\prime|^2}{\|{\bf U}\|^2 \, \|{\bf U}^\prime\|^2}.
\end{equation}
Values close to one indicate that the modal shapes ${\bf U}, {\bf U}^\prime$ are highly correlated. In COMSOL, the sampled modal shapes can be batch-exported and MAC indices can be computed against a reference set of eigenshapes. 
\begin{figure}
\centering
\includegraphics[width=\linewidth,clip,trim=2cm 2cm 2cm 2cm]{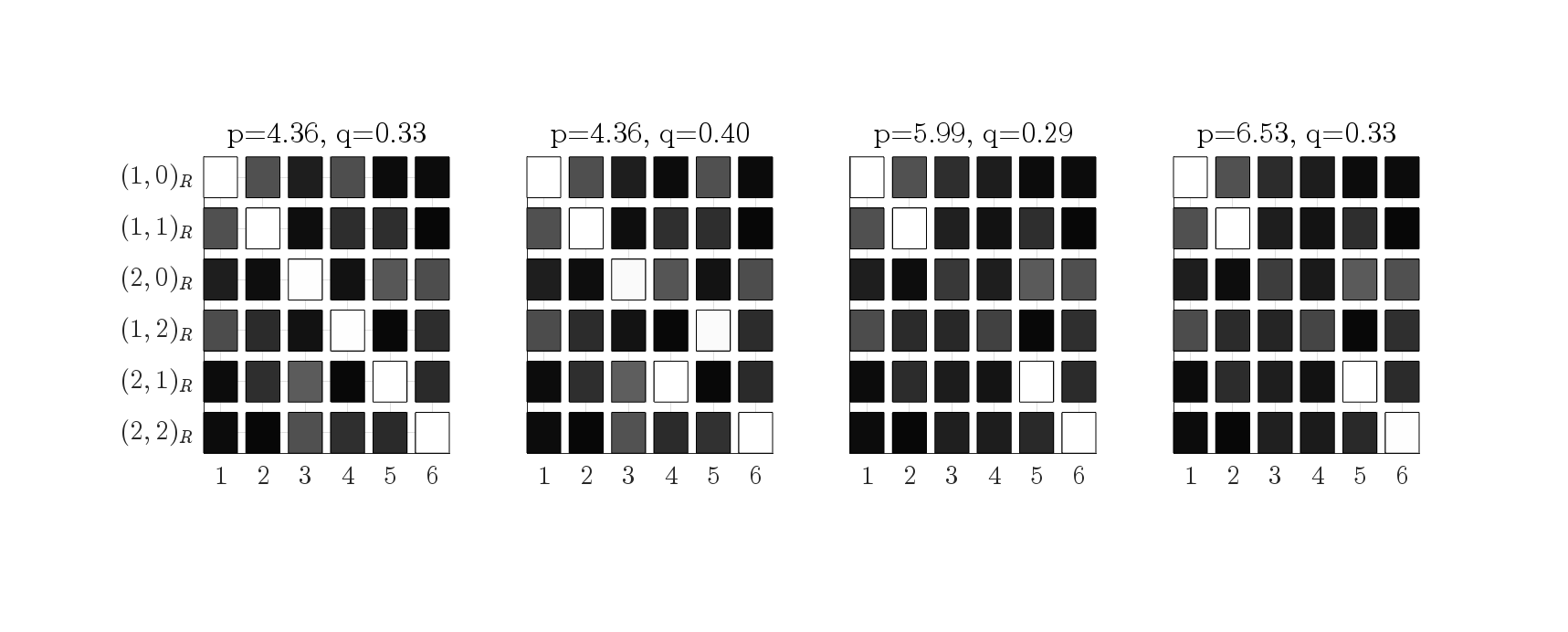}
\caption{Modal Assurance Criterion (MAC) for the plate $\sigma = 223/114$, $\mathcal{B} = \{\text{C-F-F-F}\}$, under various choices of the parameters $p,q$. A white square denotes a high correlation between the reference mode shapes (denoted by $_R$), and the batch of modal shapes for the indicated $p,q$, ordered according to increasing frequency from 1 to 6. Modal crossings may take place in some cases, such as in the second square $(p=4.36,q=0.40)$. Some ambiguity exists in identifying modes (2,0) and (1,2), as seen in the last two matrices. These cases are discarded when performing the fit. The reference set of modes is selected as the batch $p=4.36, q = 0.29$.}\label{fig:MAC}
\end{figure}
Figure \ref{fig:MAC} reports a few MAC matrices. Note that modal crossings may sometimes take place. Furthermore, in many cases out of the training batch, the modes (2,0) and (1,2) cannot be easily recognised, as one can see from the same figure. This happens because the nodal lines are bent compared to the reference $(x,y)$ axes: these modes deform considerably when changing $p,q$, not allowing a clear identification of the modal indices $(m,n)$. This is a property of the quarter-sawn plates reported in \cite[p.84]{fletcher_book_2012}. The cases for which a clear modal identification is difficult are therefore discarded. See also Figure \ref{fig:modalShapes}. 
\begin{figure}
\includegraphics[width=\linewidth,clip, trim = 1cm 1cm 1cm 0cm]{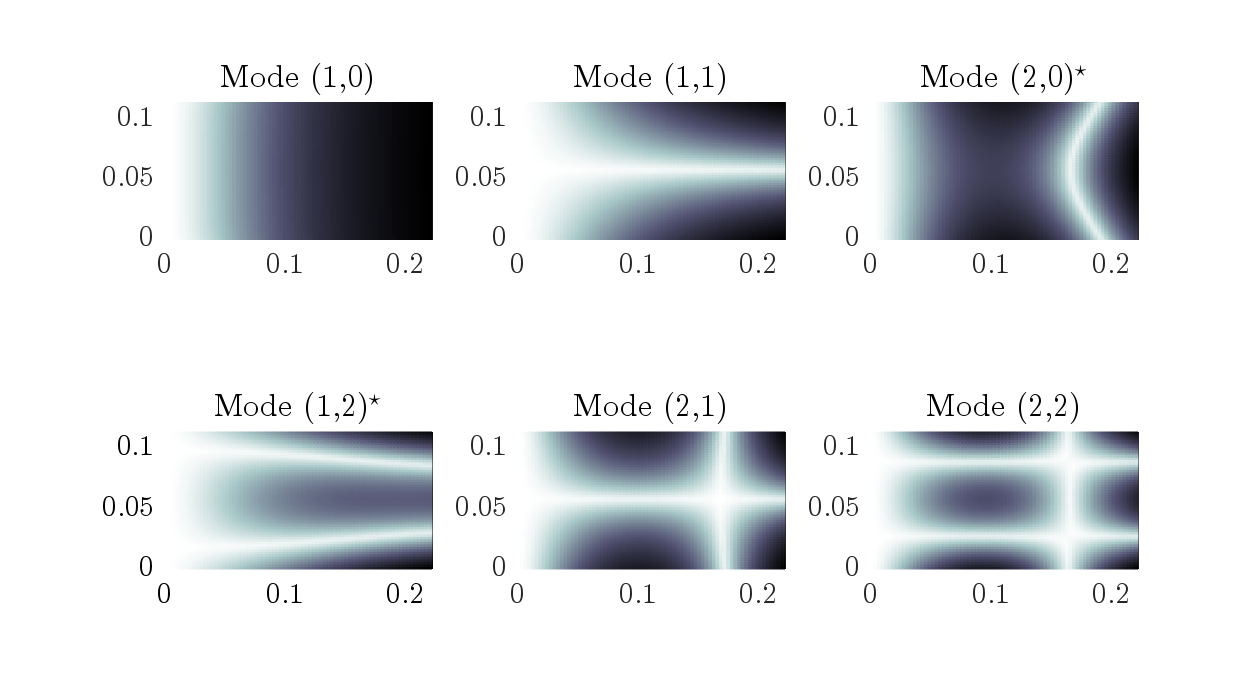}
\caption{Modal shapes for a plate with $\sigma = 223/114$, $\mathcal{B} = \{\text{C-F-F-F}\}$, $p=6.53$, $q=0.29$. Modes (1,2) and (2,0) are marked with a star as assigning unambiguous modal indices may be difficult in some cases. See also Figure \ref{fig:MAC}. }\label{fig:modalShapes}
\end{figure}

\begin{figure}
\centering
\includegraphics[width=\linewidth]{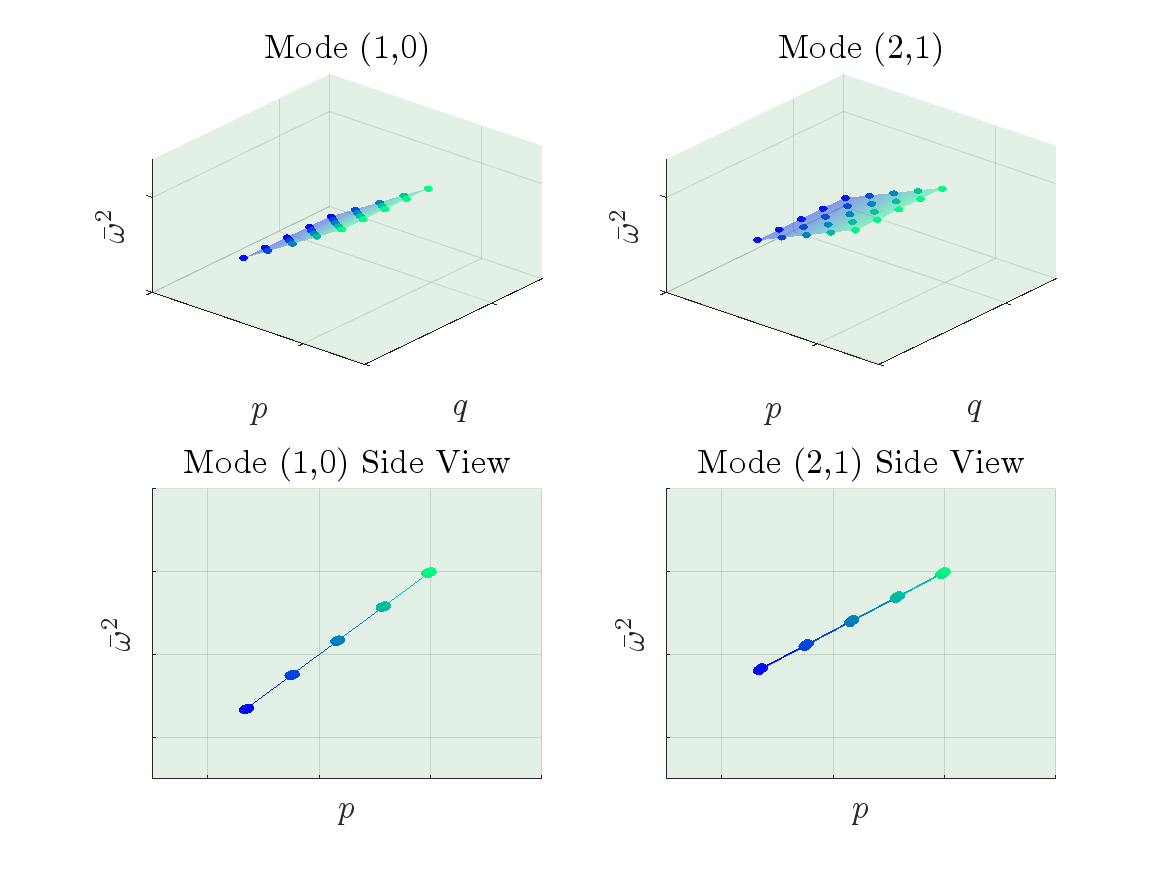}
\caption{Linear dependence of the non-dimensional frequencies $\bar\omega^2_{m,n}$ on the elastic constant ratios $p,q$. Here, the plate has $\sigma = 223/114$, and $\mathcal{B} = \{\text{C-F-F-F}\}$ (a cantilever plate clamped along the radial direction). The dots represent the non-dimensional modal frequencies as a function of $p,q$, and the planes are plotted according to \eqref{eq:linDepOm} using the values of $a_{m,n}$, $b_{m,n}$, $c_{m,n}$ listed in Table \ref{tab:LinearFitResults} .}\label{fig:LinDep}
\end{figure}

The coefficients $a_{m,n}$,  $b_{m,n}$,  $c_{m,n}$ can now be estimated in Matlab via a constrained linear fit, such that $a_{m,n}$,  $b_{m,n}$,  $c_{m,n}$ are non-negative, as per \eqref{eq:linDepOm}. The results are summarised in Table \ref{tab:LinearFitResults}. Note that the coefficients of determination $(R^2)$ are very close to one in all cases, highlighting the ability of the linear relationship \eqref{eq:linDepOm} to reproduce the correct modal frequency as a function of $p,q$. A visual representation of the fit results for the modes is given in Figure \ref{fig:LinDep}. 
\begin{table}
\addtolength{\tabcolsep}{5pt}
\centering
\begin{tabular}{lcccccc} 
 &   (1,0) & (1,1) & (2,0) & (1,2) & (2,1) & (2,2)  \\ 
    \midrule
$a_{m,n}$ & 3.21 &  3.99 & 117.0 &   14.4 & 124 &  134  \\
$b_{m,n}$ & 0.00 & 0.94 &  41.9 & 1720 &  15.6 & 1910  \\
$c_{m,n}$ & 0.00 &  40.5 &  13.6 &  155 & 328 & 1260  \\
    \midrule 
$R^2$ &  1.00 &   1.00 & 0.999 &  0.999 &   1.00 &    1.00 \\
\end{tabular}
\caption{Linear fit results for a plate with $\sigma = 223/114$, and $\mathcal{B} = \{\text{C-F-F-F}\}$ (a cantilever plate clamped along the $y$ (radial) direction). The coefficients of determination $R^2$ are also listed. }\label{tab:LinearFitResults}
\end{table}

\subsection{Assessment of the method in the case $\mathcal{B} = \{\text{C-F-F-F}\}$} \label{sec:NumericalCantilever}

The modal coefficients from Table \ref{tab:LinearFitResults} can be used to estimate the rigidity constants of a batch of test numerical plates sharing the same boundary conditions $\mathcal{B} = \{\text{C-F-F-F}\}$ and aspect ratio  $\sigma = 223/114$. It is helpful to test the methodology for plates presenting $p,q$ outside of the training ranges to assess the predicting power of the fit. Hence, three test plates are considered, all sharing the same parameters as per Table \ref{tab:TrainingPlateParams}, but selecting $E_x$, $E_y$, so as to place $p,q$ inside or outside their respective training ranges. Table \ref{tab:TestPlateParameters} collects all such parameters. 
\begin{table}
\addtolength{\tabcolsep}{-0.5pt}
\centering
\begin{tabular}{lcc|cc|ccc} 
 &   $E_x/E_x^0$ & $E_y/E_y^0$ & $p$ & $q$ & $\text{err}_x (\%)$ &  $\text{err}_y(\%)$ &  $\text{err}_{s}(\%)$  \\ 
    \midrule
Plate 1 & 1.3 & 0.7 &  7.1  &  0.25  & 0.77 & 0.9 & 0.78  \\   
Plate 2 & 1.0 & 0.7 &  5.4 & 0.25 & 0.65 & 1.2 & 0.031  \\
Plate 3 & 1.3 & 1.0 &  7.1 & 0.36 & 0.28 & 0.11 & 0.57  \\
\end{tabular}
\caption{Test plate parameters and errors for the case $\sigma = 223/114$, $\mathcal{B} = \{\text{C-F-F-F}\}$. The other plate parameters are as per Table \ref{tab:TrainingPlateParams}. The training ranges, from Section \ref{sec:LinDepCheck}, are: $4.3 < p < 6.6$, $0.28 < q < 0.45$. The errors are defined as: $\text{err}_x = 100(D_x-d_x)/D_x$, $\text{err}_y = 100(D_y-d_y)/D_y$, $\text{err}_{s} = 100(D_{s}-d_{s})/D_{s}$. The modes considered in this test are (1,0),(1,1),(2,1),(2,2). The values of $E^0_x$, $E^0_y$ are given in Table \ref{tab:TrainingPlateParams}.}\label{tab:TestPlateParameters}
\end{table}
For all test plates, the eigenfrequencies are computed numerically, and the rigidity constants $d_x$, $d_y$, $d_{s}$ are estimated according to \eqref{eq:dEstim}, using the values of the modal coefficients ${\bf a}, {\bf b}, {\bf c}$ as in Table \ref{tab:LinearFitResults}. 

The results are reported in Table \ref{tab:TestPlateParameters}, displaying errors below 1.2\%. It is worth commenting on the results. An important aspect concerns the selection of an appropriate batch of modes for the estimation of the elastic constants. Modes presenting an influence on all three constants may be used, provided that at least three of them present linearly independent modal coefficients $a_{m,n}, b_{m,n}, c_{m,n}$, as is the case for the coefficients listed in Table \ref{tab:LinearFitResults}. From the same table, it is clear that using only modes (1,0), (1,1) and (2,1) will not yield reliable results, as $D_y$ has no influence on these modes' dynamics  (the $b_{n,m}$ coefficients are either zero or small compared to $a_{n,m}$, $c_{n,m}$). In other words, the matrix in the least-square formula \eqref{eq:dEstim} becomes poorly conditioned \cite{wei2007condition}.

\subsubsection{Sensitivity analysis for the case $\mathcal{B} = \{\text{C-F-F-F}\}$ }

The numerical benchmark test carried out above highlights the ability of the proposed method to recover the correct elasticity constants with very small errors. However, the ``experimental'' frequencies were obtained via controlled numerical simulation and are unaffected by all sources of error besides a small inaccuracy due to the numerical approximation. Furthermore, the dynamics of real plates may be somewhat more complicated than what is implied by model \eqref{eq:DimEqCnt}: perfectly orthotropic conditions are seldom encountered in wooden boards; thickness profiles may not be completely uniform; the boundary conditions, especially of fixed type, may not be implemented exactly. Whilst care must be taken to ensure that the experimental plate and the measurement setup are as close to the ideal case as possible, experimental frequencies are likely affected by some form of error. Thus, it is important to test the reliability of the method using perturbed frequencies. 
\begin{table}
\addtolength{\tabcolsep}{-2.0pt}
\centering
\begin{tabular}{lcccccc|ccc} 
 &   $\epsilon_{(1,0)} $ &  $\epsilon_{(1,1)}$ &  $\epsilon_{(2,0)}$ &  $\epsilon_{(1,2)}$ & $\epsilon_{(2,1)}$ &  $\epsilon_{(2,2)}$ & $\text{err}_x(\%) $ &  $\text{err}_y(\%) $ &  $\text{err}_s(\%) $  \\ 
    \midrule
Test 1 & 1.8 & -1.7 & 1.1 &-0.73 & 1.6 & 0.95 & -4.2 & 13.0 & -3.8   \\
Test 2 &-1.6 & 0.76 & 1.6 & -1.1 & -1.4 & -2.0 & -3.5 & 7.2 & 11.0 \\
Test 3 & 0.03 & -0.1 & 1.6 & -0.95 & 1.3 & -1.4 & -6.4 & 11.0 & 8.6  \\
Test 4 & 0.76 & 0.89 & 0.66 & 0.79 & 1.7 & 1.6 & -3.4 & 9.2 & -6.0 \\
Test 5 & -0.94 & -0.55 & 0.026 & -0.35 & 0.15 & -0.92 & -2.4 & 9.1 & 4.3    \\
\end{tabular}
\caption{Sensitivity test for the Plate 1 of Section \ref{sec:NumericalCantilever}, Table \ref{tab:TestPlateParameters}, with $\sigma = 223/114$, $\mathcal{B}=\{\text{C-F-F-F}\}$. The perturbation $\epsilon$ was selected randomly within the interval $\pm 2\%$ with respect to the exact modal eigenfrequency, and is reported in the table as a percentage. The definitions of $\text{err}_x$, $\text{err}_y$ and $\text{err}_s$ are as per Table \ref{tab:TestPlateParameters}.}\label{tab:SensitivityCFFF}
\end{table}
Results are reported in Table \ref{tab:SensitivityCFFF}, where the rigidity constants of Plate 1 from Table \ref{tab:TestPlateParameters} are computed using a perturbed set of eigenfrequencies, with perturbations generated randomly within $\pm 2\%$ of the original eigenfrequencies. As expected, the errors are higher, though still in line with error bounds reported with previous methods, such as via the FEMU-3DVF \cite{viala_COMP_2018}.

\subsection{Assessment of the method in the case $\mathcal{B} = \{\text{C-C-C-C}\}$} \label{sec:NumericalClamped}

As a second case, consider the fully-clamped case with $\sigma = 150/103$. First, the modal coefficients $a_{m,n}$, $b_{m,n}$, $c_{m,n}$ are computed as explained in Section \ref{sec:LinDepCheck}. 
\begin{table}
\addtolength{\tabcolsep}{5pt}
\centering
\begin{tabular}{lcccccc} 
 &   (0,0) & (0,1) & (1,0) & (0,2) & (1,1) & (1,2)  \\ 
    \midrule
$a_{m,n}$ & 236 & 237.5 & 1790 & 242  & 1792 & 1799  \\
$b_{m,n}$ & 1214 & 8589 & 1652  & 31910  & 10180 & 35250  \\
$c_{m,n}$ & 146.9 & 552.3 & 534.9  & 1196 & 2051 & 4469 \\
    \midrule 
    $R^2$ & 1.00 & 1.00 & 1.00 & 1.00 & 1.00 & 1.00 \\
\end{tabular}
\caption{Linear fit results for a plate with $\sigma = 150/103$, and $\mathcal{B} = \{\text{C-C-C-C}\}$. The coefficients of determination $R^2$ are also listed. }\label{tab:LinearFitResultsCC}
\end{table}
\begin{figure}
\includegraphics[width=\linewidth, clip, trim = 1cm 0cm 1cm 0cm]{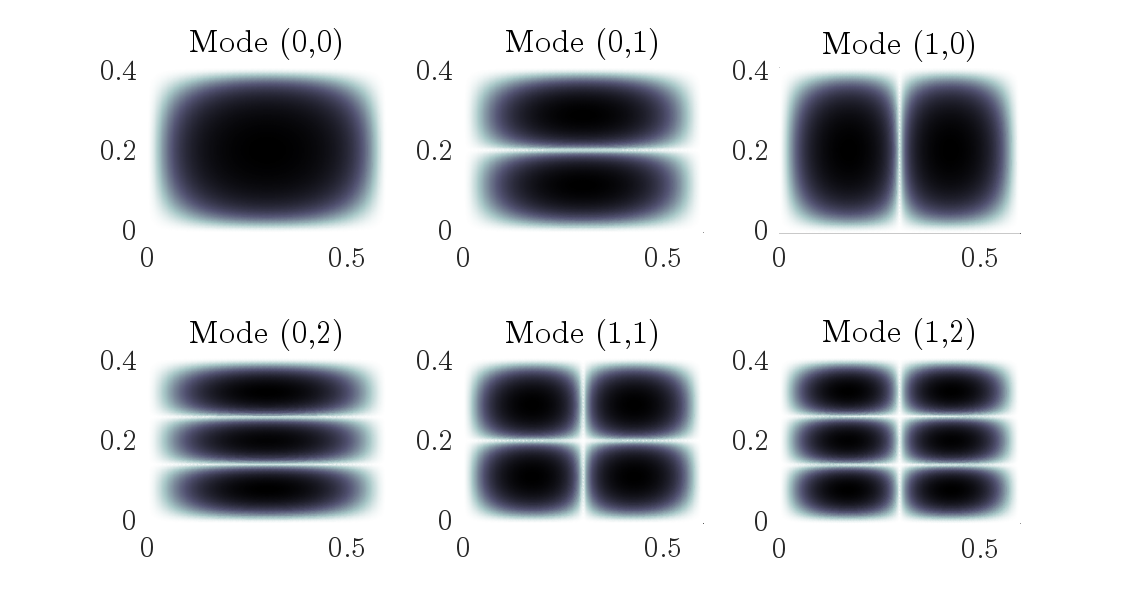}
\caption{Modal shapes for a plate with $\sigma = 150/103$, $\mathcal{B} = \{\text{C-C-C-C}\}$, $p=6.53$, $q=0.29$. }\label{fig:CCCCmodes}
\end{figure}
These are reported in Table \ref{tab:LinearFitResultsCC}. The coefficients of determination ($R^2$) are all equal to 1.00 in this case, highlighting the validity of the linearity assumption \eqref{eq:linDepOm} in the fully-clamped case. The modal shapes for the same modes are reported in Figure \ref{fig:CCCCmodes}.
\begin{table}
\addtolength{\tabcolsep}{-0.5pt}
\centering
\begin{tabular}{lcc|cc|ccc} 
 &   $E_x/E_x^0$ & $E_y/E_y^0$ & $p$ & $q$ & $\text{err}_x (\%)$ &  $\text{err}_y(\%)$ &  $\text{err}_{s}(\%)$  \\ 
    \midrule
Plate 1 & 1.3 & 0.7 &  7.1 & 0.25 & 0.81 & 0.86 & 0.12 \\   
Plate 2 & 1.0 & 0.7 &  5.4 & 0.25 & -0.28 & -0.26 & 0.043 \\
Plate 3 & 1.3 & 1.0 &  7.1 & 0.36 &- 0.12 & -0.12 & 0.036 \\
\end{tabular}
\caption{Test plate parameters and errors for the case $\sigma = 150/103$, $\mathcal{B} = \{\text{C-C-C-C}\}$.  The training ranges, from Section \ref{sec:LinDepCheck}, are: $4.3 < p < 6.6$, $0.28 < q < 0.45$. The errors are defined as: $\text{err}_x = 100(D_x-d_x)/D_x$, $\text{err}_y = 100(D_y-d_y)/D_y$, $\text{err}_{s} = 100(D_{s}-d_{s})/D_{s}$. All six modes are considered in this test.}\label{tab:TestPlateCCCC}
\end{table}
A numerical benchmark test, similar to what was done for the previous case, is reported in Table \ref{tab:TestPlateCCCC}. The elastic constants are estimated with errors below 0.9\%. A sensitivity analysis, analogous to what was done previously, is performed on the fully clamped plate. The results, reported in Table \ref{tab:SensitivityCCCC}, show that $D_x$, $D_y$ are estimated consistently well across all tests, and within small error bounds. A somewhat larger error is observed in the estimation of $D_s$. In order to account for such a larger sensitivity, a statistical approach can be adopted, as will be shown in Section \ref{sec:Discussion}.
\begin{table}
\addtolength{\tabcolsep}{-1.0pt}
\centering
\begin{tabular}{lcccccc|ccc} 
 &   $\epsilon_{(0,0)} $ &  $\epsilon_{(0,1)}$ &  $\epsilon_{(1,0)}$ &  $\epsilon_{(0,2)}$ & $\epsilon_{(1,1)}$ &  $\epsilon_{(1,2)}$ & $\text{err}_x(\%) $ &  $\text{err}_y(\%) $ &  $\text{err}_s(\%) $  \\ 
    \midrule
Test 1 & 0.64 & 0.51 & 0.37 & -1.0 & 1.6 & -0.95 & -2.5 & 3.7 & 10.6   \\
Test 2 & 2.0 & -1.0 & -1.7 & -0.86 & -1.4 & 1.7 & 8.3 & 8.4 & -52 \\
Test 3 & 1.2 & 1.4 & 2.0 & -1.9 & -0.65 & 0.63 & -1.1 & 5.7 & -6.9  \\
Test 4 & -0.49 & -1.2 & -1.1 & 0.8 & 0.98 & 1.3 & 3.7 & 3.3 & -30.0 \\
Test 5 & 0.89 & -0.95 & -0.2 & -1.9 & 0.75 & -1.6 & -1.1 & 5.4  & 11.0    \\
\end{tabular}
\caption{Sensitivity test for the Plate 1 of Table \ref{tab:TestPlateCCCC}, with $\sigma = 150/103$, $\mathcal{B}=\{\text{C-C-C-C}\}$. The perturbation $\epsilon$ was selected randomly within the interval $\pm 2\%$ with respect to the exact modal eigenfrequency, and is reported in the table as a percentage. The definitions of $\text{err}_x$, $\text{err}_y$ and $\text{err}_s$ are as per Table \ref{tab:TestPlateCCCC}.}\label{tab:SensitivityCCCC}
\end{table}

\subsection{Interpretation of the results in terms of the Moore–Penrose inverse}
\begin{figure}
\includegraphics[width=\linewidth,clip,trim=1cm 3cm 1cm 2cm]{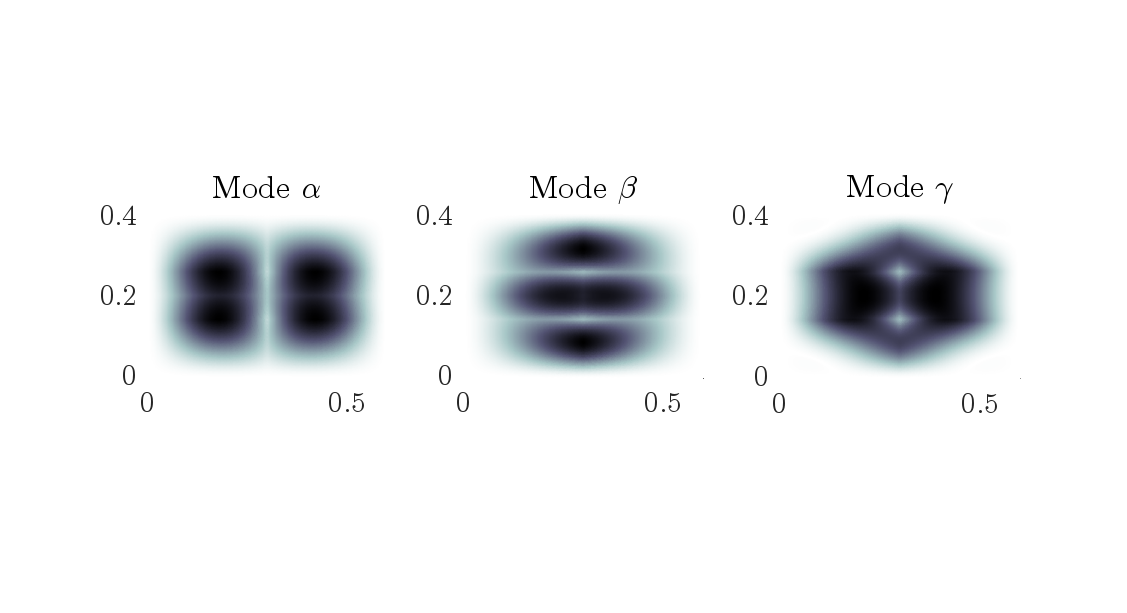}
\caption{Combined mode shapes for the fully-clamped plate. These are obtained by combining the mode shapes of Figure \ref{fig:CCCCmodes} using the Moore–Penrose inverse ${\bf X}^\dagger$, so to obtain the modes $\alpha,\beta,\gamma$ for which the dependence on, respectively, $E_x$, $E_y$ and $G_{xy}$ is completely decoupled.}\label{fig:CombinedModes}
\end{figure}

In the classic methods by McIntyre and Woodhouse and by Caldersmith and Freeman \cite{mcintyre_AM_1988,caldersmith_JCA_1990}, it is possible to extract the elasticity constants easily since these appear decoupled in two ``beam'' modes and one ``shear'' mode of the fully-free plate, $\mathcal{B} = \{\text{F-F-F-F}\}$. No such triplet of modes exists for different boundary conditions, particularly of fixed type where the effects of the elastic constants are densely coupled for all the modes, as one can see easily from Table \ref{tab:LinearFitResultsCC}.
Yet, the proposed methodology is able to retrieve the correct elastic constant values with very small errors. The linear formulation of the method offers a viable interpretation of such results. 
From \eqref{eq:dEstim}, one has that the rigidity constants are obtained in terms of the Moore-Penrose inverse ${\bf X}^\dagger$, given by  \cite{barata2012moore}:  
\begin{equation}
{\bf X}^\dagger := ({\bf X}^\intercal {\bf X} )^{-1} {\bf X}^\intercal.
\end{equation}
This linear operation reduces the overdetermined system comprising $N > 3$ equations to a $3 \times 3$ system, obtained as a linear combination of the $N$ available modes with the shortest Euclidian distance between the experimental and reconstructed eigenfrequencies. It is natural to combine the eigenshapes via ${\bf X}^\dagger$ and to observe the resulting modal shapes. This operation is unnecessary to determine the elastic constants as such, though it helps develop an intuitive interpretation of the results. The operation is a simple matrix multiplication:
\begin{equation}\label{eq:MoorePenrose}
\begin{bmatrix}
{\boldsymbol \alpha}^\intercal \\
{\boldsymbol \beta}^\intercal \\
{\boldsymbol \gamma}^\intercal
\end{bmatrix} = 
{\bf X}^\dagger\begin{bmatrix}
{\bf U}_{1}^\intercal \\
... \\
{\bf U}_{N}^\intercal
\end{bmatrix}
,
\end{equation}
where ${\boldsymbol \alpha}$, ${\boldsymbol \beta}$, ${\boldsymbol \gamma}$ are the combined mode shape column vectors, and where ${\bf U}_{i}$ is the $i^{th}$ normalised column eigenshape.  A visual representation is offered in Figure \ref{fig:CombinedModes}, where the effects of $E_x$, $E_y$ and $G_{xy}$ are now decoupled for modes $\alpha,\beta,\gamma$, and are clearly visible. 
The possibility of combining modes in the current framework is one of the most interesting findings of this work allowing the implementation of the method using various choices of boundary conditions. The next section presents two such experiments.

\section{Experimental Setup}\label{sec:experiments}
In order to validate the presented methodology, two experimental plates are now considered, with the same boundary conditions and aspect ratios as the numerical plates presented in Section \ref{sec:NumBench}. They are representative of soundboards used in the production of stringed musical instruments: a kantele and an acoustic guitar. 

\subsection{Tonewood specimen with $\mathcal{B} = \{\text{C-F-F-F}\}$} \label{ExperimentalCFFF}

A sample of Finnish tonewood spruce was obtained from a master luthier based in Finland. The sample was a leftover coming from the production of a concert kantele, a Finnish plucked stringed musical instrument \cite{tahvanainen2023concert}. This board has dimensions $L_x^0 = 25.3$ cm, $L_y=11.4$ cm, $L^0_z = 5$ mm. The thickness is reduced to $L_z = 3$ mm using a belt sander, improving the thin-plate approximation.

\begin{figure}
\centering
    \includegraphics[width = 0.8\linewidth,clip,trim=3cm 1.5cm 3cm 1cm]{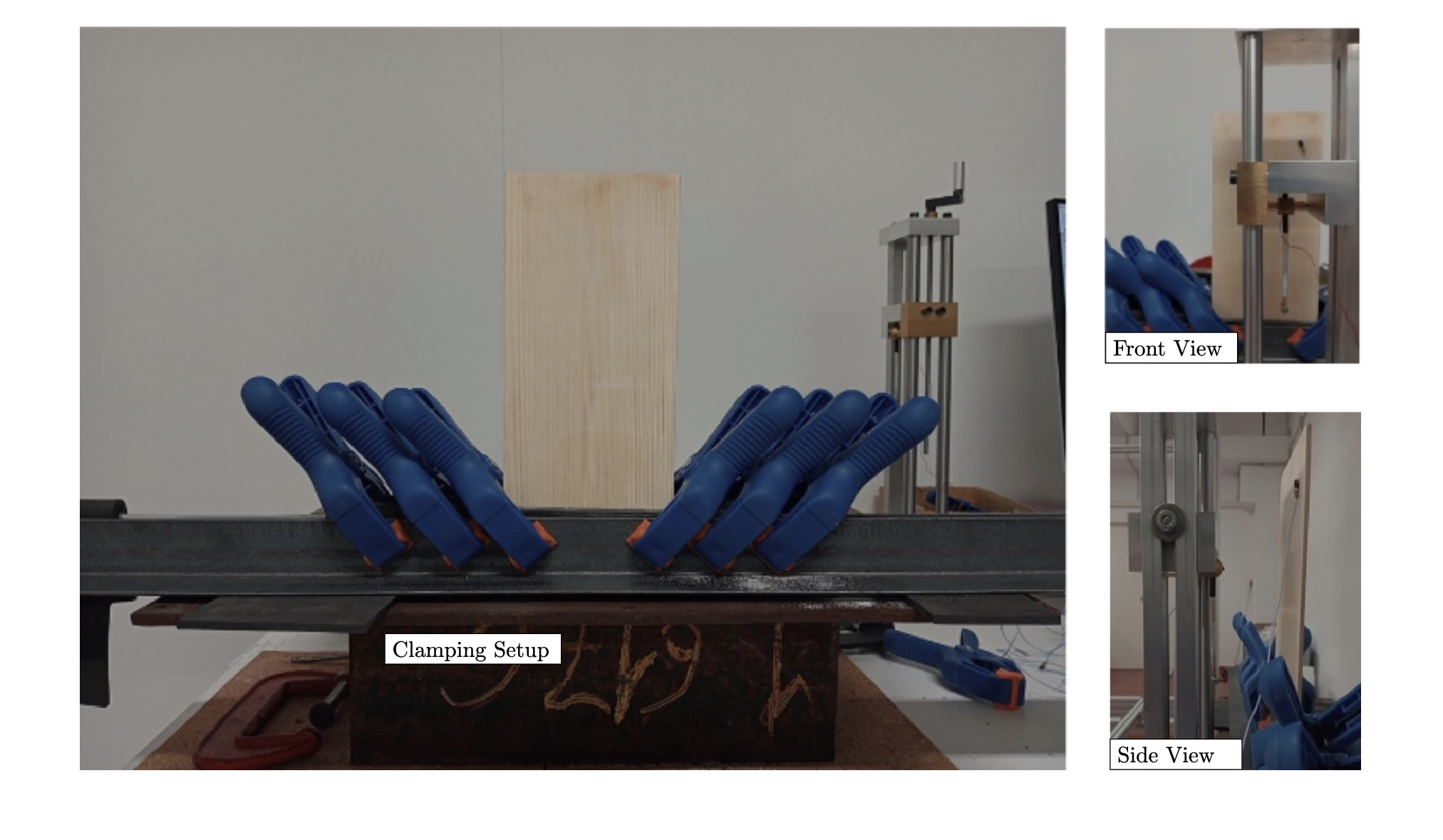}
\caption{Experimental setup for the tonewood sample with $\mathcal{B} = \{\text{C-F-F-F}\}$, $\sigma = 223/114$, for which the modal coefficients are listed in Table \ref{tab:LinearFitResults}. The hammer is mounted on a pendulum which stops after recoiling to avoid double hits.} \label{fig:ClampingSystem} 
\end{figure}

To realise the experimental boundary conditions, a clamping system is implemented as in a previous study \cite{duran2023experimentally}. Two angular 1.8 kg iron elements are clamped together by using a total of six spring plastic clamps distributed along the entire length of the two angular components. To better distribute the pressure along the $y$  axis, an additional flat 1.2 kg iron plate is placed between one of the two angular elements and the specimen under test while an additional layer of rubber material is wrapped around the clamped plate's side to prevent unwanted plate rotations. Figure \ref{fig:ClampingSystem} shows the final experimental setup for the plate boundary conditions. 
The clamping mechanism shortens the specimen along the $x$ direction, such that the effective length results as $L_x = 22.3$ cm. The final plate dimensions $L_x$, $L_y$, $L_z$ as well as the density $\rho$, are the same as the ones reported in Table \ref{tab:TrainingPlateParams}. In particular, the modal coefficients $a_{m.n}$, $b_{m,n}$, $c_{m.n}$ for this plate are listed in Table \ref{tab:LinearFitResults}.

In this experiment, a miniature impact hammer (PCB 086E80) is used to excite the plate and a monoaxial accelerometer (PCB 352A21) is used to measure the plate's impulse response. A vinyl cover is used on the hammer's hard tip to reduce the damage caused by the impacts on the wooden plate surface. 
This softer tip limits the force spectrum of the excitation signal \cite{ewins2009modal}, though it still delivers sufficient energy to excite the board in the frequency range of interest (below 1 kHz). To conduct the measurements, the impact hammer roves across six equally spaced measurement positions whereas the accelerometer is mounted and kept still in the same position during the entire measurement campaign. To account for the measurement uncertainty caused by the low reproducibility of the impact hammer force signal \cite{avitabile2017modal}, five different impulse responses are averaged over each measurement position. Particular care is taken to avoid placing the accelerometer close to the nodal lines of the interested modes listed in Table \ref{tab:LinearFitResults}. Finally, the collected data is post-processed in Matlab to analyse the frequency response functions from each measurement position.

\begin{figure}
    \centering
    \includegraphics[width = \linewidth, clip, trim = 1cm 0.5cm 1cm 0.2cm]{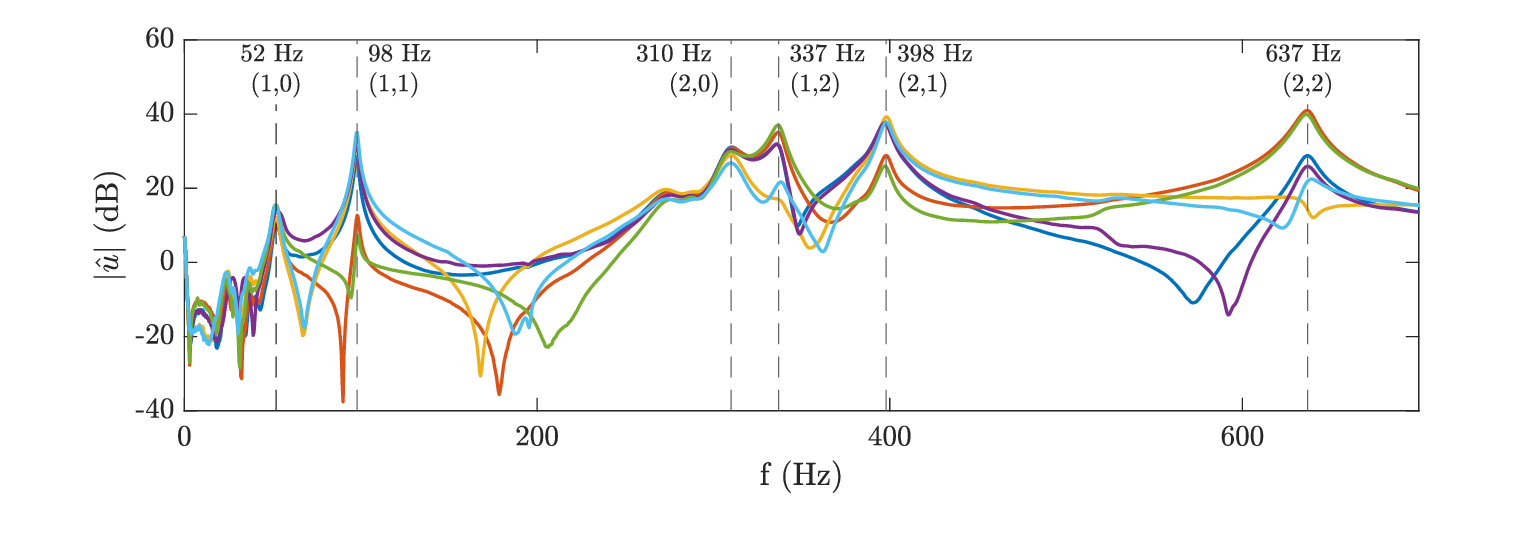}
\caption{Frequency spectra obtained from the six measurement points assessed on the tonewood spruce plate. The frequency range of interest here is up to 800 Hz. Dashed lines mark the spatially averaged experimental frequencies.}\label{fig:PlateSpectra} 
\end{figure}

\begin{figure} 
\centering
\includegraphics[width= 0.8\linewidth]{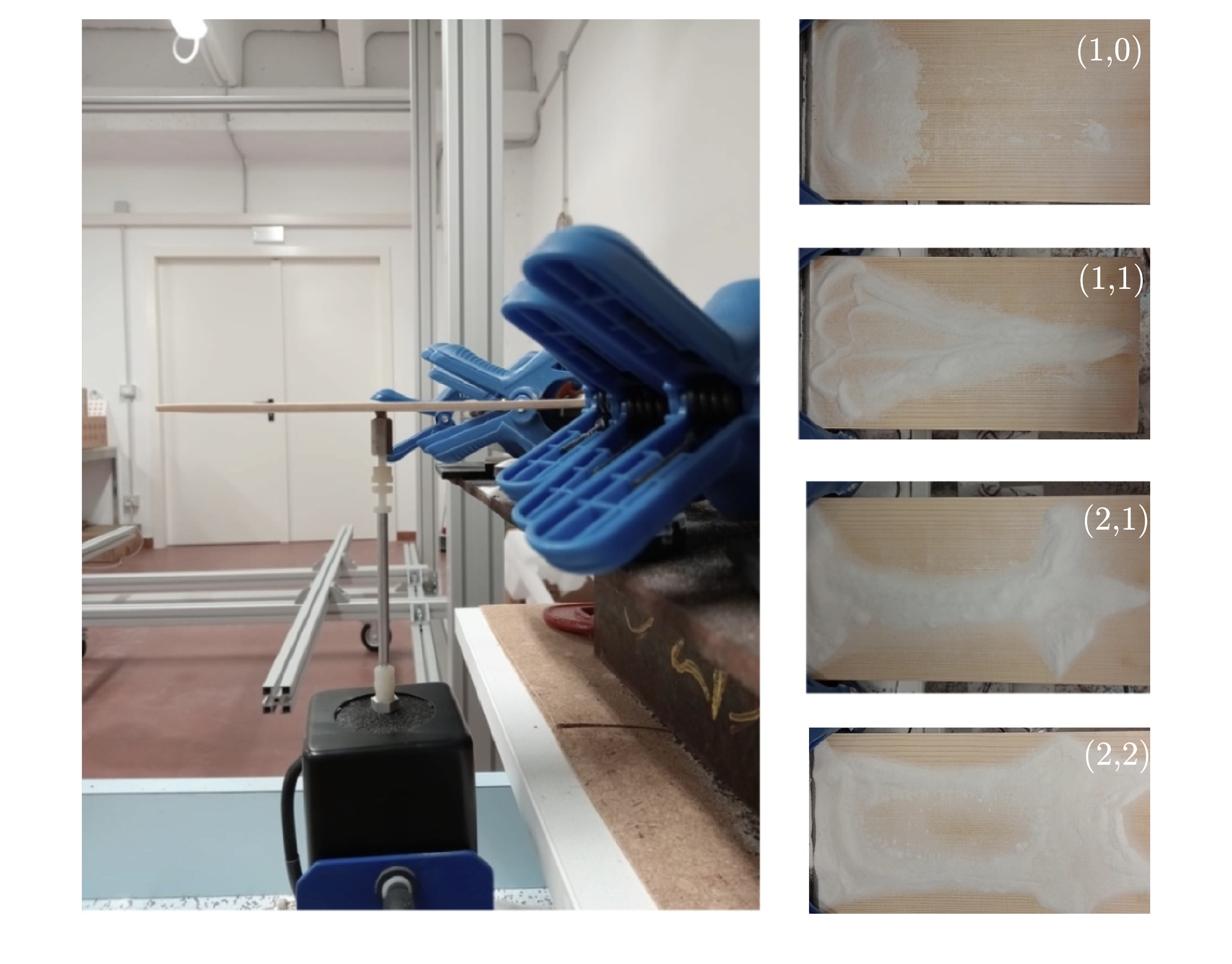}
\caption{Experimental setup including the miniature shaker (left) and Chladni patterns (right) on the tonewood sample. }\label{fig:ChladniSetup}
\end{figure}

 Figure \ref{fig:PlateSpectra} 
shows the resulting frequency spectra and Table \ref{tab:AvgPlatesFreqs} 
summarises the experimental eigenfrequencies identified through a ``peak finding'' routine implemented in Matlab and averaged across all the measurement positions. 
\begin{table}
\renewcommand{\arraystretch}{1.5}
\centering
\begin{tabular}{ccccccc} 
& (1,0) & (1,1) & (2,0) & (1,2) & (2,1) & (2,2) \\
\cline{2-7} 
    {\footnotesize Measured freqs. (Hz)} & 52  & 98  & 310  & 337  & 398 & 637\\ 
    \bottomrule
    \end{tabular}
\caption{Spatially averaged experimental frequencies and identified mode shapes for the cantilever Finnish spruce tonewood. }\label{tab:AvgPlatesFreqs}
\end{table}
As mentioned in Section \ref{sec:Methods}, a crucial step of the presented methodology requires the identification of the nodal lines along the $x$ and $y$ directions of the tonewood's fibres, in order to associate the measured frequencies with the corresponding modal coefficients $a_{m,n}, b_{m,n}, c_{m,n}$. The modal shapes can be simply observed through the well-known technique by Chladni \cite{chladni2015treatise}. Here, a mini shaker (PCB 2004E) is used to excite the cantilever plate which is now horizontally oriented to facilitate the mounting of the shaker to the plate surface (as shown in Figure \ref{fig:ChladniSetup}) through the use of adhesive material (i.e. beeswax). Pure tones having the same frequencies as the six measured eigenfrequencies reported in Table \ref{tab:AvgPlatesFreqs} are used as excitation signals for the plate covered with off-the-shelf white sand. Figure \ref{fig:ChladniSetup} shows the observed Chladni patterns. 

\subsection{Acoustic guitar top board with $\mathcal{B} = \{\text{C-C-C-C}\}$} 

\begin{figure} 
\centering
\includegraphics[width = \linewidth]{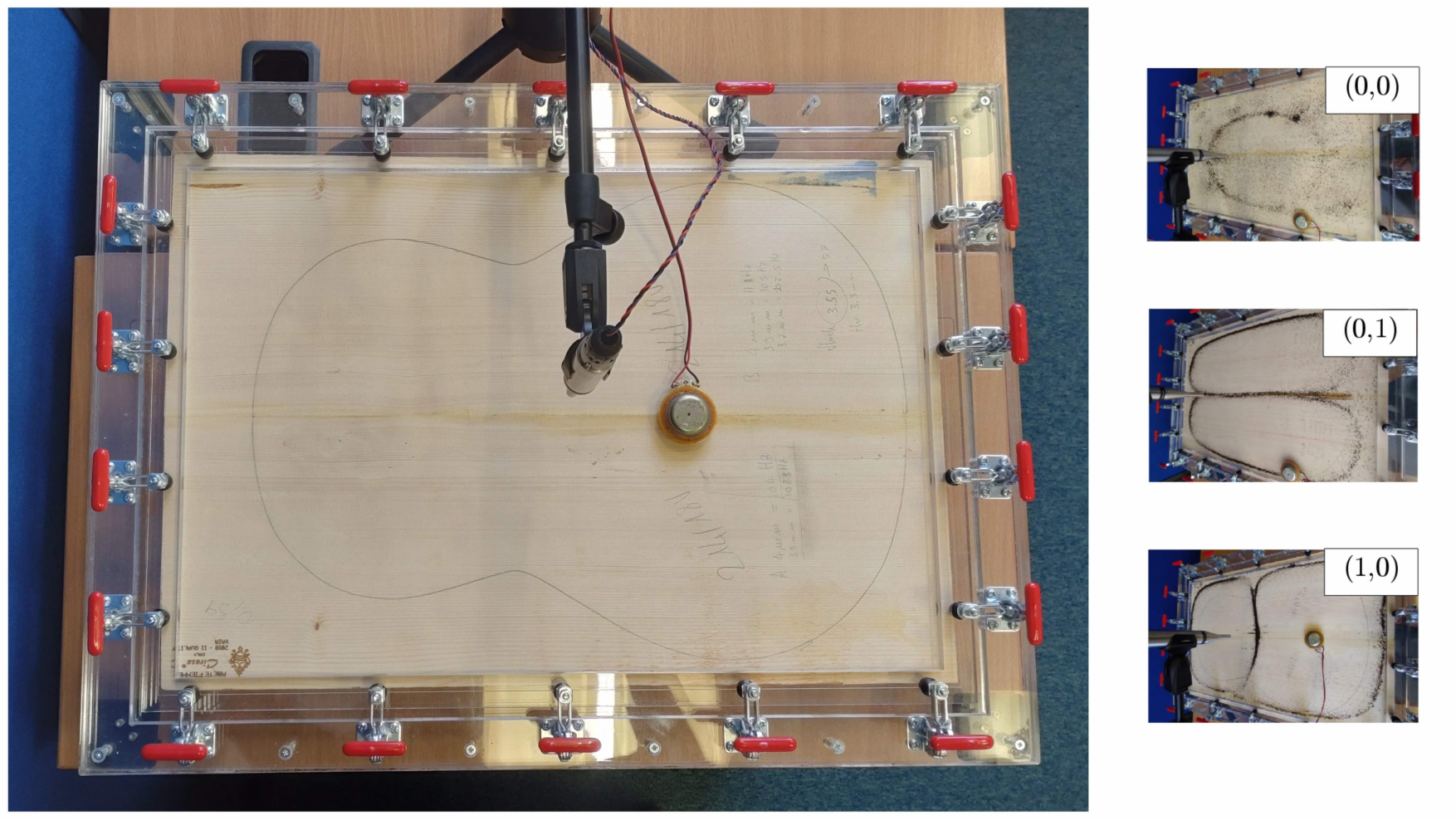}
\caption{Experimental setup and Chladni patterns for the Fiemme Valley red spruce guitar top.}\label{fig:LudoMeasSetup}
\end{figure}

In a second experiment, a full guitar board is considered. 
This is the assembly of two rectangular quarter-sawn spruce billets, referred to as a \emph{book-matched top}. Glueing together the half-boards to form a full guitar board represents a preliminary step in the construction and design of guitars conducted by luthiers and builders \cite{rau_JASA_2021}. The board was purchased from Ciresa\footnote{\url{https://www.ciresafiemme.it/}}, a Fiemme Valley red spruce reseller, and is here tested under fully clamped boundary conditions, mimicking those occurring in the final assembled instrument \cite{gore2011contemporary}. To do so, an 8 kg plexiglass frame is built and used to clamp the board along all its edges. Here, equally distributed clamps are mounted along the frame's edges and are used to apply pressure on an additional plexiglass counter-frame which is resting directly on the board under test. The experimental setup is shown in Figure \ref{fig:LudoMeasSetup}. The clamped board's dimensions are: $L_x = 60$ cm,  $L_y = 41.2$ cm, $L_z = 3$ mm, and its density is $\rho = 399$ kg $\cdot$ m$^{-3}$. Thus, the aspect ratio is $\sigma = 105/103$, and the modal coefficients $a_{m,n}, b_{m,n}, c_{m,n}$ are as per Table \ref{tab:LinearFitResultsCC}.

Given the larger size and weight of the guitar board compared to the cantilever plate illustrated in Section \ref{ExperimentalCFFF}, a higher force amplitude is needed to excite all the interested modes of vibration with sufficient energy. Furthermore, as the experimental fully clamped constraints require the board to be now horizontally fixed to a frame, a different experimental setup is here employed for ease of operation \cite{ausiello2018guitar, ausiello2021quantitative}. Accordingly, the guitar board is excited by an 8 $\Omega$ nominal impedance 48 g electro-dynamic exciter with a 25 mm diameter voice coil. Such a setup allows for a higher amplitude of excitation and ease of operation on the horizontal board.

An exponential sine sweep ranging from 45 Hz to 8 kHz is used as the test signal. Similarly to the previous measurement campaign, the exciter roves across four different measurement points distributed over the guitar's board. As a receiver, an Earthworks MD30 class-1 microphone is placed at a 125 mm distance from the board to capture the output signal from the guitar board. Finally, the recorded data is post-processed using the Aurora plugins \cite{farina2007advancements}, yielding the impulse responses which are then further processed and analysed in Matlab. Table \ref{tab:LudoFreqs} reports the measured eigenfrequencies.
\begin{table}
\renewcommand{\arraystretch}{1.5}
\centering
\begin{tabular}{lcccccc} 
& (0,0) & (0,1) & (1,0) & (1,1) & (0,2) & (1,2) \\
\cline{2-7} 
    {\footnotesize Measured freqs. (Hz)} & 57  & 100  & 130  & 162  & 170  & 224 \\ 
    \bottomrule
    \end{tabular}
\caption{Spatially averaged experimental frequencies and identified mode shapes for the fully clamped red spruce plate. }\label{tab:LudoFreqs}
\end{table}
For the presented case study, Chladni patterns are obtained by exciting the plate at the measured frequencies using pure tones and dried tea leaves, and are visible in Figure \ref{fig:LudoMeasSetup}.

\section{Results and Discussion}\label{sec:Discussion}
\begin{figure}
\includegraphics[width=\linewidth,clip,trim=2.0cm 0.0cm 2.5cm 0.2cm]{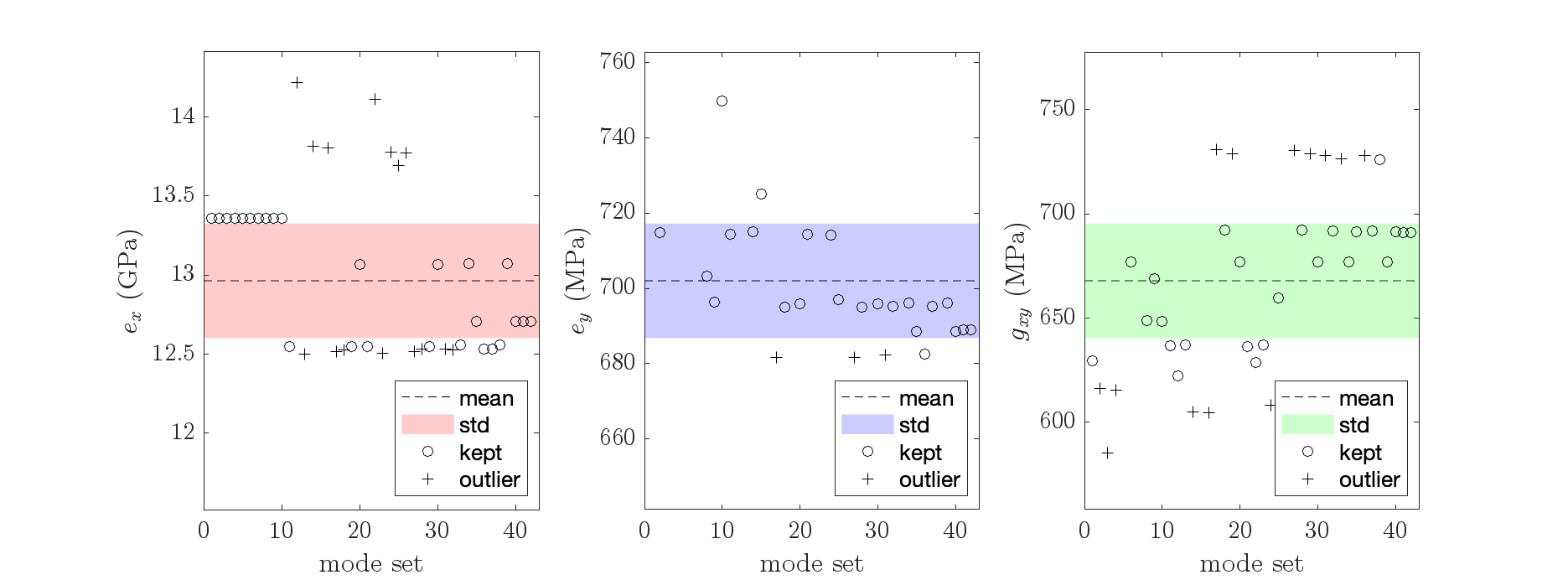}
\caption{Elastic constant estimates for the Finnish spruce tonewood. Forty-two mode sets are considered, corresponding to all possible combinations containing at least three experimental modes out of the six measured ones. Mean values and standard deviations are represented as dashed lines and coloured bands, respectively. Outliers are selected as belonging to the twentieth percentile away from the mean.}\label{fig:ResultsFinnish}
\end{figure}

\begin{figure}
\includegraphics[width=\linewidth,clip,trim=2.0cm 0.0cm 2.5cm 0.2cm]{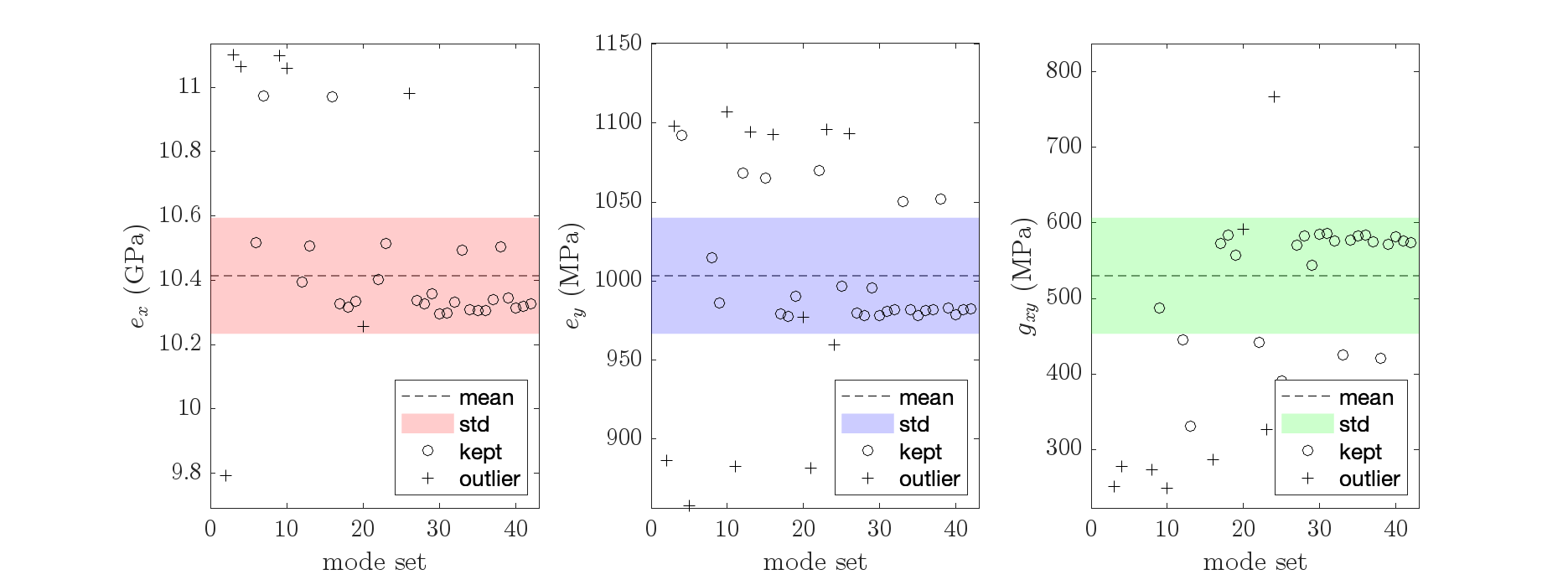}
\caption{Elastic constant estimates for the Fiemme Valley red spruce tonewood. The same analysis as per Figure \ref{fig:ResultsFinnish} holds.}\label{fig:ResultsFiemme}
\end{figure}

The elastic constants for the Finnish and the Fiemme Valley spruce tonewoods are retrieved using the experimental frequencies from Tables \ref{tab:AvgPlatesFreqs} and \ref{tab:LudoFreqs} respectively, the modal coefficients from Tables \ref{tab:LinearFitResults} and \ref{tab:LinearFitResultsCC}, and formulae \eqref{eq:dEstim} and \eqref{eq:EstElCnt}. Various combinations of modes can be used here. Both cases include six identified experimental modes; thus, it is possible to run \eqref{eq:dEstim} using forty-two combinations containing at least three modes out of six. This allows performing many estimates of the elastic constants, yielding a statistically significant set from which mean values and standard deviations can be computed. It is best to exclude from the statistics all spurious estimates. First, negative estimates should be excluded (these appear in some cases when the Moore-Penrose is poorly conditioned \cite{wei2007condition}, as discussed previously). Second, outliers should be excluded. These may be given in terms of deviation from the mean: here, all estimates lying in the twentieth percentile away from the mean are excluded. Results are summarised in Figures \ref{fig:ResultsFinnish} and \ref{fig:ResultsFiemme}.

The mean values and standard deviations of $e_x$, $e_y$, $g_{xy}$ for the two plates are summarised in Table \ref{tab:results}. Note that the standard deviations are very low in all cases (below 5\%), except for the shear modulus in the fully clamped case. This is a consequence of the sensitivity of the method under fully-clamped conditions, as discussed in Section \ref{sec:NumericalClamped}, though the statistical approach adopted here is able to contain the deviation to within a reasonable threshold. 
The estimated mean values are  in line with previously reported values for spruce, such as in \cite[p.96]{bucur_book_2016}.
\begin{table}
\renewcommand{\arraystretch}{1.5}
\centering
\begin{tabular}{llccc} 
& & $e_x$ (GPa) & $e_y$ (MPa) & $g_{xy}$ (MPa)   \\
\cline{2-5} 
\multirow{ 2}{*}{\footnotesize{Finnish spruce (cantilever)}} &  mean & 13.0 & 702 & 668 \\
& std & 2.8\% & 2.2\% & 4.1\% \\
\midrule
\multirow{ 2}{*}{\footnotesize{Fiemme Valley spruce (clamped)}} &  mean & 10.4 & 1003 & 530 \\
& std & 1.73\% & 3.7\% & 14\% \\
\bottomrule
\end{tabular}
\caption{Elastic constants mean values and standard deviations for the two spruce tonewoods. Values are in line with previously reported values for spruce, such as in \cite[p.96]{bucur_book_2016}. }\label{tab:results}
\end{table}

The average elastic constant values can be used as input parameters in COMSOL, and the numerical frequencies obtained can be assessed against the measured experimental frequencies. This is done in Table \ref{tab:AssCFFF} and  \ref{tab:AssCCCC}, where the numerical frequencies are in both cases predicted very accurately: errors fall below 2$\%$ in all cases, highlighting the accuracy of the proposed methodology in retrieving appropriate elastic constant values. 
\begin{table}
\addtolength{\tabcolsep}{-0.5pt}
\centering
\begin{tabular}{lcc|ccc} 
 & Meas. (Hz) & Num. (Hz) & $\Delta f_n$ (Hz) & $\frac{\Delta f_n}{f_n}$ (\%) & ${\Delta f_n}$ (cent)  \\ 
    \cmidrule{2-6}
(1,0)          &  52 &  51 &  -1  &  -1.5 & -25     \\   
(1,1)          &  98 &  100 &  2 &  1.9  & 32     \\
(2,0)          & 311 & 315 &      4 &   1.3  &22     \\
(1,2)          & 337 & 336 &      -1 & -0.4 &-7     \\
(2,1)          & 398 & 394 &-4  & -0.9 &-16     \\
(2,2)          & 637 & 629 & -8 & -1.2  & -21
\end{tabular}
\caption{Errors between the experimentally measured and numerically computed eigenfrequencies, for the Finnish spruce plate, under cantilever boundary conditions. The numerical frequencies were computed in COMSOL using the mean elastic constant values from Table \ref{tab:results}. }\label{tab:AssCFFF}
\end{table}

\begin{table}
\addtolength{\tabcolsep}{-0.5pt}
\centering
\begin{tabular}{lcc|ccc} 
 & Meas. (Hz) & Num. (Hz) & $\Delta f_n$ (Hz) & $\frac{\Delta f_n}{f_n}$ (\%) & ${\Delta f_n}$ (cent)  \\ 
    \cmidrule{2-6}
(0,0)            & 57 &  56 & -1 &  -1.8 & -31     \\   
(0,1)            & 100 &  98 &-2 &  -2 & -35 \\
(1,0)            & 130 & 130 &   0 &      0 &    0 \\
(1,1)            & 162 & 161 & -1 & -0.6 & -11\\
(0,2)            & 170 & 169 & -1  &-0.6 & -10 \\
(1,2)            & 224 & 221 & -3 &   -1.3 & -23
\end{tabular}
\caption{Errors between the experimentally measured and numerically computed eigenfrequencies, for the Fiemme Valley red spruce plate, under fully clamped boundary conditions. The numerical frequencies were computed in COMSOL using the mean elastic constant values from Table \ref{tab:results}.}\label{tab:AssCCCC}
\end{table}

\section{Conclusion}

In this work, a method to estimate the elastic constants of thin, rectangular orthotropic panels was offered, as an application of non-destructive inverse parameter estimation methods. Compared to previously available techniques in the literature, the proposed framework does not rely on a specific set of boundary conditions or modes, allowing multiple estimates of the elastic constants via simple matrix inverses. This way, not only can mean values be estimated, but also deviations. The enabling idea is the identification of a linear relationship between the elastic constants and the square of the non-dimensional plate frequencies, with modal-dependent coefficients that can be easily computed from a batch of numerical training plates. An interpretation of the method was offered in terms of the Moore-Penrose inverse, which is used to combine the eigenmodes to yield three modes in which the influence of the thin-plate elastic constants is completely decoupled. A number of numerical and experimental tests showed the reliability of the proposed methodology in cases of interest in musical acoustics. 

\section*{Acknowledgements}

This work received funding from the European Research Council (ERC) under the Horizon2020 framework, grant number 950084 - StG - NEMUS. Francesco Vai and Roberto Budini from the Department of Industrial Engineering at the University of Bologna are kindly acknowledged for designing and building the impact hammer pendulum. Ciresa is kindly acknowledged for assisting in the selection of the Fiemme Valley tonewood. Master luthier Pekka Lovikka is thanked for providing the Finnish tonewood samples.






\end{document}